\documentclass[12pt]{article}
\usepackage{ametsoc}
\usepackage{amsmath} 
\usepackage{amssymb} 
\usepackage{amsfonts}
\usepackage{graphicx,color}                 
\usepackage{subfigure}
\usepackage{lineno}

\usepackage[absolute,overlay]{textpos}

\newcommand{\myabstract}{

}
\begin{document}
%
%

\title{\textbf{\large{Ensemble Kalman filtering with residual nudging: an extension to state estimation problems with nonlinear observation operators}}}
%
%
\author{\textsc{Xiaodong Luo}
				\thanks{\textit{Corresponding author address:}
				International Research Institute of Stavanger (IRIS), Thorm{\o}hlens Gate 55, 5008 Bergen, Norway
				\newline{E-mail: xiaodong.luo@iris.no}} \\
				\textit{\footnotesize{International Research Institute of Stavanger (IRIS), 5008 Bergen, Norway}}	
\and
\centerline{\textsc{Ibrahim Hoteit}}\\
\centerline{\textit{\footnotesize{King Abdullah University of Science and Technology (KAUST), Thuwal 23955-6900, Saudi Arabia}}}							
}
%
\ifthenelse{\boolean{dc}}
{
\twocolumn[
\begin{@twocolumnfalse}
\amstitle

\begin{center}
\begin{minipage}{13.0cm}
\begin{abstract}
	\myabstract
	\newline
	\begin{center}
		\rule{38mm}{0.2mm}
	\end{center}
\end{abstract}
\end{minipage}
\end{center}
\end{@twocolumnfalse}
]
}
{
\amstitle

\begin{abstract}
\myabstract 
The ensemble Kalman filter (EnKF) is an efficient algorithm for many data assimilation problems. In certain circumstances, however, divergence of the EnKF might be spotted. In previous studies, the authors proposed an observation-space-based strategy, called residual nudging, to improve the stability of the EnKF when dealing with linear observation operators. The main idea behind residual nudging is to monitor and, if necessary, adjust the distances (misfits) between the real observations and the simulated ones of the state estimates, in the hope that by doing so one may be able to obtain better estimation accuracy. 

In the present study, residual nudging is extended and modified in order to handle nonlinear observation operators. Such extension and modification result in an iterative filtering framework that, under suitable conditions, is able to achieve the objective of residual nudging for data assimilation problems with nonlinear observation operators. The 40 dimensional Lorenz 96 model is used to illustrate the performance of the iterative filter. Numerical results show that, while a normal EnKF may diverge with nonlinear observation operators, the proposed iterative filter remains stable and leads to reasonable estimation accuracy under various experimental settings. 
\end{abstract}
}

\section{Introduction}
The ensemble Kalman filter (EnKF, see \citealp{Burgers-analysis,Evensen2006}) and its variants (including, for example, \citealp{Anderson-ensemble,Bishop-adaptive,Hoteit2002,Luo-ensemble,Pham2001,Tippett-ensemble,Wang-which,Whitaker-ensemble}) can be considered as Monte Carlo implementations of the celebrated Kalman filter \citep{Kalman-new}, in the sense that the mean and covariance of the Kalman filter are evaluated based on a finite (often small) number of samples of the underlying model states. Because of its ability to handle large-scale data assimilation problems, and its relative simplicity in implementation, the EnKF has received great attention from researchers in various fields. 

In data assimilation, there are certain factors that may influence the performance of the EnKF. For instance, if the EnKF is implemented with a relatively small ensemble size, then the filter will often be subject to sampling errors. This may lead to some adverse effects (especially in high-dimensional models), including, for instance, under-estimation of the variances of state variables, over-estimation of the correlations between different state variables, and rank deficiency of the sample error covariance matrix \citep{Whitaker-ensemble,Hamill2009}. In the literature, it is customary to adopt two auxiliary techniques, called covariance inflation \citep{Anderson-Monte} and covariance localization \citep{Hamill-distance}, to improve the performance of the EnKF. Intuitively, covariance inflation compensates for the under-estimated variances by artificially increasing it to some extent. It also increases the robustness of the EnKF from the point of view of $H_{\infty}$ filtering theory \citep{Luo2011_EnLHF}. Various methods of covariance inflation are proposed in the literature, for example, see \citet{Altaf2013-improving,Anderson-Monte,Anderson2007,Anderson2009,Bocquet2011-ensemble,Bocquet2012-combining,Luo2011_EnLHF,Luo2013-covariance,Miyoshi2011-Gaussian,meng2007tests,Ott-local,song2013adjoint,Triantafyllou2012-assessing,Whitaker2012-evaluating,zhang2004impacts}. On the other hand, covariance localization aims to taper the over-estimated correlations through, for instance, a Schur product between the sample error covariance matrix and a certain tapering matrix. In effect, this also increases the rank of the sample error covariance matrix \citep{Hamill2009}.  

Even equipped with both covariance inflation and localization, the EnKF may still suffer from filter divergence in certain circumstances, especially when there is substantial uncertainty, e.g., in terms of model and/or observation errors, in data assimilation problems (see, for example, the numerical results in \citealp{Luo2012-residual}). To mitigate filter divergence, in previous studies \citep{Luo2014-efficient,Luo2013-covariance,Luo2012-residual} we considered a strategy, called data assimilation with residual nudging (DARN), which monitors and, if necessary, adjusts the distances (called residual norms) between the real observations and the simulated ones. Our numerical results showed that, under certain circumstances, a data assimilation algorithm equipped with residual nudging is not only more stable against filter divergence, but also performs better in terms of estimation accuracy.   

The analytical and numerical results in \citet{Luo2014-efficient,Luo2013-covariance,Luo2012-residual} also show that, for linear observation operators, one is able to control the magnitudes of the residual norms under suitable conditions. An issue that we did not address yet is the nonlinearity in the observation operators. Our main motivation here is thus to fill this gap. To this end, we recast DARN as a least-squares problem and adopt an iterative filtering framework\footnote{Here, by ``iterative'' we mean the presence of an iteration process (Eq. (\ref{eq:iteration_formula})) in each data assimilation cycle.} to tackle the nonlinearity in the observation operators. Using this iterative filtering framework, one can achieve the objective of residual nudging under suitable conditions. For convenience, we refer to the observations from a linear (or nonlinear) observation operator as ``linear observations'' (or ``nonlinear observations''), when it causes no confusion.     

This work is organized as follows. Section \ref{sec:rn_linear} introduces the idea of DARN and outlines the method used in \citet{Luo2013-covariance} for residual nudging with linear observations. In Section \ref{sec:rn_nonlinear}, the aforementioned method is extended and modified to tackle problems with nonlinear observations. In Section \ref{sec:experiments}, various experiments are conducted to compare the proposed method with some existing algorithms in the literature. In addition, the stability of the proposed method is also investigated under different experimental settings. Finally, Section \ref{sec:conclusion} concludes the work.

\section{Residual nudging with linear observations} \label{sec:rn_linear}

In this work we consider the state estimation problem in the following systems 
\begin{linenomath*}
\begin{subequations} \label{eq:ps}
\begin{align}
\label{subeq:dynamical}	&\mathbf{x}_{k} = \mathcal{M}_{k,k-1} ( \mathbf{x}_{k-1}  ) + \mathbf{u}_k \, , \\
\label{subeq:observation}	&\mathbf{y}_{k} = \mathcal{H}_{k} ( \mathbf{x}_{k}  ) + \mathbf{v}_k \, , 
\end{align}
\end{subequations}
\end{linenomath*}
where $\mathbf{x}_{k}$ is an $m$-dimensional model state and $\mathbf{y}_{k}$ is the corresponding $p$-dimensional observation; $\mathcal{M}_{k,k-1}$ is the model transition operator that maps the model state $\mathbf{x}_{k-1}$ at time instant $(k-1)$ to the next time instant $k$, and $\mathcal{H}_{k}$ is the observation operator that projects the model state $\mathbf{x}_{k}$ onto the observation space; and $\mathbf{u}_k \in \mathbb{R}^{m}$ and $\mathbf{v}_k \in \mathbb{R}^{p}$ are the model and observation errors, respectively. We assume that the observation error $\mathbf{v}_k$ has zero mean and a non-singular covariance matrix $\mathbf{R}_k$. In the discussion below, the time index $k$ is often uninvolved and thus dropped for ease of notation. 

In this section, we focus on the case with linear observations. To this end, we rewrite the observation operator $\mathcal{H}$ as $\mathbf{H}$. Suppose that $\mathbf{y}^o$ is the real observation at a certain time instant, and $\hat{\mathbf{x}}$ is the estimate of the underlying model state $\mathbf{x}$. Then the \textit{residual} $\hat{\mathbf{r}}$ is defined as the difference between the simulated observation $\mathbf{H} \, \hat{\mathbf{x}}$ and the real value $\mathbf{y}^o$, i.e., $\hat{\mathbf{r}} = \mathbf{H} \, \hat{\mathbf{x}} - \mathbf{y}^o$.   

Given a vector $\mathbf{z} \in \mathbb{R}^{p}$ in the observation space, we use the weighted Euclidean norm
\begin{linenomath*}
\begin{equation} \label{eq:l2_norm}
\Vert \mathbf{z} \Vert_{\mathbf{R}} \equiv \sqrt{\mathbf{z}^T \, \mathbf{R}^{-1} \, \mathbf{z}} 
\end{equation}
\end{linenomath*}
to measure its length. Under this setting, the objective of residual nudging is to find an analysis estimate $\hat{\mathbf{x}}^a$, such that the weighted Euclidean norm of the corresponding analysis residual $\hat{\mathbf{r}}^a$ (the analysis residual norm hereafter) satisfies $\Vert \hat{\mathbf{r}}^a \Vert_{\mathbf{R}} \leq  \beta_u \sqrt{p}$ for a pre-chosen scalar $\beta_u$ ($\beta_u > 0$). Readers are referred to \citet{Luo2014-efficient,Luo2013-covariance,Luo2012-residual} for the rationale behind this choice. To prevent over-fitting the observation, it may also be desirable to let $\beta_l \sqrt{p} \leq \Vert \hat{\mathbf{r}}^a \Vert_{\mathbf{R}}$ for another scalar $\beta_l$ ($0 < \beta_l < \beta_u$). Combining these constraints, the objective thus becomes
\begin{linenomath*}
\begin{equation} \label{eq:residual_norm_target_ineq}
\beta_l \sqrt{p} \leq \Vert \hat{\mathbf{r}}^a \Vert_{\mathbf{R}} \leq  \beta_u \sqrt{p} \, .
\end{equation}
\end{linenomath*} 

Two methods were proposed in \citet{Luo2014-efficient,Luo2013-covariance,Luo2012-residual} for the purpose of residual nudging. In \citet{Luo2014-efficient,Luo2012-residual} it was suggested to solve a linear equation first, and then combine the resulting solution (called the ``observation inversion'') with the original state estimate. In a follow-up work \citep{Luo2013-covariance}, residual nudging was recast as a problem of choosing a proper covariance inflation factor, and some sufficient conditions in this regard were explicitly derived for the analysis residual norm to be bounded in the interval $[\beta_l \sqrt{p}, \, \beta_u \sqrt{p}]$. In the next section, the second method will be extended and modified to handle nonlinear observations. For this reason, in what follows, we summarize the method developed in \citet{Luo2013-covariance}. Readers are referred to \citet{Luo2014-efficient,Luo2012-residual} for more details about the first method. 

In \citet{Luo2013-covariance}, a family of mean update formulae, in the form of
\begin{linenomath*}
\begin{subequations} \label{eq:mean_update_general}
\begin{align}
\label{sub_eq: mean_update}& \hat{\mathbf{x}}^a = \hat{\mathbf{x}}^b + \mathbf{G} \left(\mathbf{y}^o - \mathbf{H} \hat{\mathbf{x}}^b  \right) \, , \\
\label{sub_eq: gain_mtx} & \mathbf{G} = \hat{\mathbf{C}}^b \mathbf{H}^T \left( \delta \, \mathbf{H} \hat{\mathbf{C}}^b \, \mathbf{H}^T + \gamma \, \mathbf{R} \right)^{-1} \, ,
\end{align}
\end{subequations}
\end{linenomath*}
is considered, where $\hat{\mathbf{x}}^b$ is the sample mean of the background ensemble, $\mathbf{G}$ is a gain matrix that may be considered as a slight generalization of the Kalman gain matrix in the EnKF. The coefficients $\delta$ and $\gamma$ in $\mathbf{G}$ are some positive scalars whose ranges need to be determined in order for the analysis residual norm to be bounded in the desired interval (\ref{eq:residual_norm_target_ineq}), and $\hat{\mathbf{C}}^b$ is a symmetric, positive semi-definite matrix. In general, $\hat{\mathbf{C}}^b$ may be related, but not necessarily identical, to the sample error covariance matrix $\hat{\mathbf{P}}^b$ of the background ensemble (to be further discussed below). 

While the objective (\ref{eq:residual_norm_target_ineq}) can be achieved with the general gain matrix $\mathbf{G}$ defined in Eq. (\ref{sub_eq: gain_mtx}) \citep{Luo2013-covariance}, it suffices to consider the conventional choice of $\delta =1$ here. In this case, the gain matrix $\mathbf{G}$ resembles the Kalman gain in the EnKF, with $1/\gamma$ being analogous to the multiplicative covariance inflation factor used in \citet{Anderson-Monte}. With some algebra, it can be shown that the analysis residual norm satisfies \citep{Luo2013-covariance} 
\begin{linenomath*}
\begin{subequations} \label{eq:analysis_rn}
\begin{align}
\label{sub_eq:analysis_rn}& \Vert \hat{\mathbf{r}}^a \Vert_{\mathbf{R}} = \Vert \Phi \, (\mathbf{R}^{-1/2} \hat{\mathbf{r}}^b) \Vert_{2} \, , \\
\label{sub_eq:phi} & \Phi = \gamma \, \left( \mathbf{A} + \gamma  \mathbf{I}_p \right)^{-1} \, , \\
\label{sub_eq:A} & \mathbf{A} = \mathbf{R}^{-1/2}\mathbf{H} \hat{\mathbf{C}}^b \mathbf{H}^T \mathbf{R}^{-T/2} \, ,
\end{align}
\end{subequations}
\end{linenomath*}
where $\hat{\mathbf{r}}^b \equiv \mathbf{H} \hat{\mathbf{x}}^b - \mathbf{y}^o$ is the residual with respect to the background $\hat{\mathbf{x}}^b$; $\Vert \bullet \Vert_{2}$ denotes the standard Euclidean norm and satisfies $\Vert (\mathbf{R}^{-1/2} \hat{\mathbf{r}}^b) \Vert_{2} = \Vert \hat{\mathbf{r}}^b \Vert_{\mathbf{R}}$, with $\mathbf{R}^{1/2}$ being a square root matrix of $\mathbf{R}$; and $\mathbf{I}_p$ represents the $p$-dimensional identity matrix. For ease of notation, in Eq. (\ref{sub_eq:A}) we have used $\mathbf{R}^{-1/2}$ to denote the inverse of $\mathbf{R}^{1/2}$, and $\mathbf{R}^{-T/2}$ to represent the transpose of $\mathbf{R}^{-1/2}$.  

Given a matrix $\mathbf{M}$ and a vector $\mathbf{z}$ with suitable dimensions, one has the following inequalities \citep{Grcar2010-linear}
\begin{linenomath*}
\begin{subequations} \label{eq:mtx_ineq}
\begin{align}
\label{sub_eq:mtx_ineq1}& \Vert \mathbf{M} \, \mathbf{z} \Vert_2 \leq \Vert \mathbf{M} \Vert_2 \, \Vert \mathbf{z} \Vert_2 \, , \\
\label{sub_eq:mtx_ineq2} & \Vert \mathbf{M}^{-1} \Vert_2^{-1} \; \Vert \mathbf{z} \Vert_2  \leq \Vert \mathbf{M} \mathbf{z} \Vert_2 \, . 
\end{align}
\end{subequations}
\end{linenomath*}   
Applying (\ref{eq:mtx_ineq}) to the right-hand side of Eq. (\ref{sub_eq:analysis_rn}) with $\mathbf{M} = \Phi$ and $\mathbf{z} = (\mathbf{R}^{-1/2} \hat{\mathbf{r}}^b)$, one obtains the following sufficient conditions:
\begin{linenomath*}
\begin{subequations} \label{eq:sufficient_conditions}
\begin{align}
\label{sub_eq:sf1}& \dfrac{\xi_l}{1 - \xi_l} \, \lambda_{max}  \leq \gamma \leq \dfrac{\xi_u}{1 - \xi_u} \, \lambda_{min} \, , \\
\label{sub_eq:sf2} & \text{subject to~} \beta_l \leq \dfrac{\beta_u}{\kappa + (1-\kappa) \, \xi_u} \, ,   
\end{align}
\end{subequations}
\end{linenomath*}   
where $\xi_l = \beta_l \sqrt{p} / \Vert \hat{\mathbf{r}}^b \Vert_{\mathbf{R}}$ and $\xi_u = \beta_u \sqrt{p} / \Vert \hat{\mathbf{r}}^b \Vert_{\mathbf{R}}$, $\lambda_{max}$ and $\lambda_{min}$ are the maximum and minimum eigenvalues, respectively, of the matrix $\mathbf{A}$ in Eq. (\ref{sub_eq:A}), and $ \kappa = \lambda_{max} / \lambda_{min}$ is the corresponding condition number. In case the observation size is large, such that it is expensive to evaluate the eigenvalues of $\mathbf{A}$ directly, some alternative sufficient conditions can be obtained at a cheaper computational cost. This is omitted here for brevity and readers are referred to \citet{Luo2013-covariance} for details. Finally, we note that, from Eq. (\ref{eq:sufficient_conditions}), if only residual nudging is in consideration, then there is no restriction that $\hat{\mathbf{C}}^b$ needs to be related to the sample error covariance $\hat{\mathbf{P}}^b$. 

\textbf{Remark:} From the above deduction, one can relate residual nudging to certain forms of covariance inflation. As discussed in \citet{Luo2011_EnLHF}, a Kalman filter (or ensemble Kalman filter) with covariance inflation is essentially a $H_{\infty}$ filter (or its ensemble implementation, see \citealp{Luo2011_EnLHF}). Compared with the Kalman filter (or its ensemble variants), the $H_{\infty}$ filter (or its ensemble variants) puts more emphasis on the robustness of the estimation \citep{Simon2006}. For more details of the similarities and differences between the Kalman and $H_{\infty}$ filtering methods, readers are referred to \citet{Luo2011_EnLHF} and the references therein.
    
\section{Residual nudging with nonlinear observations} \label{sec:rn_nonlinear}
When the observation operator $\mathcal{H}$ is nonlinear, residual nudging becomes more complicated, since an explicit relation between the analysis and background residuals, similar to that in Eq. (\ref{eq:analysis_rn}) with linear observations, may no longer be available. On the other hand, if the state space of the dynamical model and the observation operator $\mathcal{H}$ are continuous, and there is at least one model state $\mathbf{x}^{\dagger}$ satisfying $\mathcal{H}(\mathbf{x}^{\dagger}) = \mathbf{y}^{o}$ (or more generally, satisfying $\Vert \mathcal{H}(\mathbf{x}^{\dagger}) - \mathbf{y}^{o} \Vert_{\mathbf{R}} \leq \beta_l \sqrt{p}$), and another model state whose residual norm is larger than $\beta_u \sqrt{p}$, then due to the continuity of $\mathcal{H}$, there exist model states satisfying (\ref{eq:residual_norm_target_ineq}), i.e., the objective of residual nudging is feasible. 

In data assimilation practices, one often has an initial state estimate with a relatively large residual norm, while it may be more difficult to have readily available a state estimate with a sufficiently small residual norm. Therefore, in what follows, we present an iterative framework that aims to construct a sequence of model states with gradually decreasing residual norms as the iteration index increases. If the iteration process (see Eq. (\ref{eq:iteration_formula}) later) is long enough, the residual norm may become sufficiently low such that (\ref{eq:residual_norm_target_ineq}) is satisfied.  

\subsection{Iteration process to reduce the residual norm} \label{subsec:iter_process}
To establish the iteration process, we first note that in cases of linear observations, the analysis $\hat{\mathbf{x}}^a$ in Eq. (\ref{eq:mean_update_general}) (with $\delta=1$) is actually the solution of the following linear least-squares problem 
\begin{linenomath*}
\begin{equation} \label{eq:linear_ls}
\underset{\mathbf{x}}{\operatorname{argmin}} \, \Vert \mathbf{y}^{o} - \mathbf{H} \mathbf{x}  \Vert_{\mathbf{R}}^2 + \gamma \Vert \mathbf{x} - \hat{\mathbf{x}}^b  \Vert_{\hat{\mathbf{C}}^b}^2 \, .
\end{equation}
\end{linenomath*}  
In line with this point of view, if the observation operator is nonlinear, then we aim to find a model state that solves the following nonlinear least-squares problem
\begin{linenomath*}
\begin{equation} \label{eq:nonlinear_ls}
\underset{\mathbf{x}}{\operatorname{argmin}} \, \Vert \mathbf{y}^{o} - \mathcal{H} (\mathbf{x})  \Vert_{\mathbf{R}}^2 + \gamma \Vert \mathbf{x} - \hat{\mathbf{x}}^b  \Vert_{\hat{\mathbf{C}}^b}^2 \, .
\end{equation}
\end{linenomath*}   
    
Some remarks regarding the cost function in (\ref{eq:nonlinear_ls}) are in order. Firstly, for the objective (\ref{eq:residual_norm_target_ineq}) of residual nudging, it is intuitive to use only the first term (called the data mismatch term hereafter) in (\ref{eq:nonlinear_ls}) as the cost function (see, for example, \citealp{kalnay2010accelerating}), which corresponds to the choice of $\gamma=0$ in (\ref{eq:nonlinear_ls}). In many situations, minimizing the term $\Vert \mathbf{y}^{o} - \mathcal{H} (\mathbf{x})  \Vert_{\mathbf{R}}^2$ alone corresponds to an ill-posed inverse problem, which has some notorious effects, e.g., non-uniqueness in the solutions and sensitivities of the solutions to (even tiny) observation errors (see, for example, the discussion in \citealp{Engl2000-regularization}). Therefore, in inverse problem theory and the geophysical data assimilation community, it is customary to introduce a certain regularization term, e.g., $\gamma \Vert \mathbf{x} - \hat{\mathbf{x}}^b  \Vert_{\hat{\mathbf{C}}^b}^2$ in (\ref{eq:nonlinear_ls}) ($\gamma > 0$), with which the aforementioned problems can be avoided or mitigated. The presence of such a regularization term makes the solution of (\ref{eq:nonlinear_ls}) approximately solve the minimization problem $\underset{\mathbf{x}}{\operatorname{argmin}} \, \Vert \mathbf{y}^{o} - \mathcal{H} (\mathbf{x})  \Vert_{\mathbf{R}}^2$, while the choice of $\gamma$ follows a certain rule (e.g., (\ref{eq:par_rule}) in this work).

In the literature, certain iteration processes are derived based on the cost function in (\ref{eq:nonlinear_ls}) with $\gamma = 1$. As in the maximum likelihood ensemble filter (MLEF, see \citealp{Zupanski-maximum}) and other similar iterative ensemble filters (see, for example, \citealp{Lorentzen2011-iterative,sakov2012iterative}), the rationale behind the choice of $\gamma=1$ may be largely explained from the point of view of Bayesian filtering, in the sense that the  solution of (\ref{eq:nonlinear_ls}) corresponds to the maximum a posterior (MAP) estimate, when both the model state and the observation follow certain Gaussian distributions. However, from a practical point of view, such an interpretation may be only approximately valid in many situations. This is not only because the Gaussianity assumption may be invalid in many nonlinear dynamical models, but also because in reality it is often very challenging to accurately evaluate certain statistics, e.g., the error covariance matrices, of both the model state and the observation in large-scale problems. 

With that said, in the iteration process below, we do not confine ourselves to a fixed cost function with either $\gamma = 0$ or $\gamma = 1$. Instead, we let $\gamma$ be adaptive with the iteration steps, which facilitates the gradual reduction of the residual norm of the state estimate, and is thus useful for the purpose of residual nudging. In \citet{Bocquet2012-combining}, an iteration process with essentially adaptive $\gamma$ values is also introduced by combining the original inflation method in \citet{Bocquet2011-ensemble} and the iterative EnKF in \citet{sakov2012iterative}. 
Note that in \citet{Bocquet2012-combining} and \citet{sakov2012iterative} the cost functions are constructed with respect to the observations both at the present time (the so-called EnKF-N) and ahead in time (the so-called IEnKF-N) with respect to the model states to be optimized, while in the current work, the observations and the model states to be estimated are in the same assimilation cycles.

For convenience of discussion, let $\{ \hat{\mathbf{x}}^i \}$ ($i = 0, 1,\dotsb$) be a sequence of state estimates obtained in the iteration process, with $\hat{\mathbf{x}}^0$ equal to the background mean $\hat{\mathbf{x}}^b$, and $\{ \gamma^i \}$ ($i = 0, 1,\dotsb$) a sequence of positive scalars that are associated with $\{ \hat{\mathbf{x}}^i \}$.  
In the iteration process, we need to (a) calculate $\hat{\mathbf{x}}^{i+1}$ based on $\hat{\mathbf{x}}^i$ and $\gamma^i$ at the previous iteration step, and (b) update the coefficient $\gamma^i$ to a new value $\gamma^{i+1}$.  

In this work, task (a) is undertaken by introducing a local linearization to the cost function in (\ref{eq:nonlinear_ls}) at each iteration step, following \citet[ch. 11]{Engl2000-regularization}. More precisely, this involves linearizing the nonlinear operator $\mathcal{H}$ locally around the most recent iterated estimate $\hat{\mathbf{x}}^i$, and replacing the reference point $\hat{\mathbf{x}}^b$ in the regularization term $\gamma \Vert \mathbf{x} - \hat{\mathbf{x}}^b  \Vert_{\hat{\mathbf{C}}^b}^2$ with $\hat{\mathbf{x}}^i$. The latter modification is introduced in accordance with the former one, since in order to make local linearization approximately valid, it is expected that the new state estimate $\hat{\mathbf{x}}^{i+1}$ should not be too far away from $\hat{\mathbf{x}}^i$. Using $\hat{\mathbf{x}}^i$ as the reference point means that the observation $\mathbf{y}^{o}$ is used multiple times in each data assimilation cycle. This choice may be justified by the fact that in many situations, the conventional non-iterative EnKF tends to be sub-optimal, due to, for instance, the nonlinearity in the dynamical model and/or the observation operator, the difficulties in accurately characterizing the statistics of the model and/or observation error(s), and the challenge in running the filter with a statistically sufficient ensemble size in large-scale applications. In such circumstances, assimilating the observations multiple times may help improve the filter's performance, as will be shown later (also see, for example, the numerical results in \citealp{Luo2012-residual}).  

Taking the above considerations into account, at each iteration step, we solve a (local) minimization problem, in the form of 
\begin{linenomath*}
\begin{equation} \label{eq:linearized_cost_func}
\underset{\mathbf{x}^{i+1}}{\operatorname{argmin}} \Vert \mathbf{y}^{o} -  \mathcal{H}( \hat{\mathbf{x}}^{i} ) - \mathbf{J}^{i} ( \mathbf{x}^{i+1} - \hat{\mathbf{x}}^{i} ) \Vert_{\mathbf{R}}^2 + \gamma^i \Vert \mathbf{x}^{i+1} - \hat{\mathbf{x}}^{i}  \Vert_{\hat{\mathbf{C}}^b}^2 \, ,
\end{equation}
\end{linenomath*}  
where $\mathcal{H}( \hat{\mathbf{x}}^{i} ) + \mathbf{J}^{i} ( \mathbf{x}^{i+1} - \hat{\mathbf{x}}^{i} )$ is the first order Taylor approximation of $\mathcal{H}( \mathbf{x}^{i+1} )$, with $\mathbf{J}^{i}$ being the Jacobian matrix of $\mathcal{H}$ at $\hat{\mathbf{x}}^{i}$ (the evaluation of the Jacobian matrix is discussed later). 

The solution $\hat{\mathbf{x}}^{i+1}$ of the minimization problem is given by \citep[ch. 11]{Engl2000-regularization} 
\begin{linenomath*}
\begin{subequations} \label{eq:iteration_formula}
\begin{align}
\label{subeq:iter_form} & \hat{\mathbf{x}}^{i+1} = \hat{\mathbf{x}}^{i} + \mathbf{G}^i \left(\mathbf{y}^{o} - \mathcal{H} (\hat{\mathbf{x}}^{i})  \right) \, , \\
\label{subeq:iter_gain} & \mathbf{G}^i = \hat{\mathbf{C}}^b (\mathbf{J}^{i})^T \left( \mathbf{J}^{i} \hat{\mathbf{C}}^b (\mathbf{J}^{i})^T + \gamma^i \, \mathbf{R} \right)^{-1} \, .
\end{align}
\end{subequations}
\end{linenomath*}       
Eq. (\ref{eq:iteration_formula}) is similar to the iteration formula in \citet[Eq. (3.49)]{Tarantola-inverse}  and the mean update formula in the ensemble square root filter (EnSRF, see \citealp{Tippett-ensemble}), or more precisely, the ensemble transform Kalman filter (ETKF, see \citealp{Bishop-adaptive,Wang-which}). In particular, if the observation operation $\mathcal{H}$ is linear and $\hat{\mathbf{C}}^b$ is equal to the background sample error covariance $\hat{\mathbf{P}}^b$ (either with or without localization), then Eq. (\ref{eq:iteration_formula}) is identical to the mean update formula in the ETKF with a multiplicative covariance inflation factor $1/\gamma^i$. In this sense, the mean update formula in the ETKF can be considered as a single step implementation of the iteration process in Eq. (\ref{eq:iteration_formula}). In addition, one may further generalize Eq. (\ref{eq:iteration_formula}) by introducing an additional scalar coefficient, say $\alpha$, in front of the gain matrix $\mathbf{G}^i$. Such an extension would then encompass the iteration processes of certain gradient-based optimization algorithms as special cases (see, for example, Eq. (A7) of \citealp{Zupanski-maximum}, where $\gamma^i \equiv 1$ during the iteration process). 

Intuitively, with the first order Taylor approximation, one has 
\begin{linenomath*}
\begin{equation} \label{eq:apprx_ineq}
\begin{split}
\Vert \mathcal{H}( \hat{\mathbf{x}}^{i+1} ) - \mathbf{y}^{o} \Vert_{\mathbf{R}}^2 & \approx \Vert [ \mathbf{I}_p - \mathbf{J}^{i} \hat{\mathbf{C}}^b (\mathbf{J}^{i})^T ( \mathbf{J}^{i} \hat{\mathbf{C}}^b (\mathbf{J}^{i})^T + \gamma^i \, \mathbf{R} )^{-1} ]  (\mathcal{H}( \hat{\mathbf{x}}^{i} ) - \mathbf{y}^{o}) \Vert_{\mathbf{R}}^2 \\
& \leq \Vert (\mathcal{H}( \hat{\mathbf{x}}^{i} ) - \mathbf{y}^{o}) \Vert_{\mathbf{R}}^2 \, ,
\end{split}
\end{equation}
\end{linenomath*}   
which implies that the residual norm of the estimate $\hat{\mathbf{x}}^{i+1}$ at the $(i+1)$-th iteration tends to be no larger than that of $\hat{\mathbf{x}}^{i}$ at the previous iteration. Similar to the situation with linear observations, in order for the inequality in (\ref{eq:apprx_ineq}) to be valid, there is no restriction that $\hat{\mathbf{C}}^b$ be related to the background sample error covariance $\hat{\mathbf{P}}^b$.    

From the point of view of the deterministic inverse problem theory, Eq. (\ref{eq:iteration_formula}) can also be considered as an implementation of the regularized Levenberg-Marquardt method (see, for example, \citealp[ch. 11]{Engl2000-regularization}), with the weight matrices for the data mismatch and regularization terms being $\mathbf{R}$ and $\hat{\mathbf{C}}^b$, respectively. Compared with the conventional Levenberg-Marquardt method (which is already used in the data assimilation community, see \citealp{Bocquet2012-combining}), there are a few differences in the regularized Levenberg-Marquardt method, which focuses more on the residual norm. For instance, in the regularized Levenberg-Marquardt method, it is customary to specify a parameter rule (see, for example, (\ref{eq:par_rule}) below) to ensure the (local) convergence of the residual norm. In addition, the stopping criterion of the iteration process in Eq. (\ref{eq:iteration_formula}) may normally involve a certain threshold of the residual norm, while in the conventional Levenberg-Marquardt method, the stopping criterion may instead be based on monitoring the relative change of the total cost function value (see, for example, \citealp{Bocquet2012-combining}).   

The relation between Eq. (\ref{eq:iteration_formula}) and the deterministic inverse problem theory is useful for establishing the rule in choosing the parameter $\gamma^i$. To this end, we adopt the following parameter iteration rule \citep[ch. 11]{Engl2000-regularization} 
\begin{linenomath*}
\begin{equation} \label{eq:par_rule}
\begin{split}
& \gamma^0 > 0 \, , \\
& \gamma^{i+1} = \rho^i \, \gamma^i, \text{~with~} 1/r < \rho^i < 1 \text{~for some scalar~} r > 1 \, , \\
& \underset{i \rightarrow +\infty}{\lim} \gamma^i = 0 \, ,
\end{split}
\end{equation}
\end{linenomath*}                        
in which the scalar sequence $\{\gamma^i\}$ gradually reduces to zero as $i$ tends to $+\infty$, where the presence of the lower bound $1/r$ for the coefficient $\rho^i$ aims to prevent any abrupt drop-down of $\gamma^i$ to zero. When Eq. (\ref{eq:iteration_formula}) is used in conjunction with Eq. (\ref{eq:par_rule}), it can be analytically shown that the residual norms of the sequence of state estimates $\{\hat{\mathbf{x}}^{i}\}$ converge to zero locally, 
provided that the equation $\mathcal{H}(\mathbf{x}) = \mathbf{y}^{o}$ is solvable (and some other conditions are satisfied, see, e.g., \citealp{jin2010regularized}). Of course, as discussed previously, it may not be desirable to have a too small residual norm in order to prevent over-fitting the observation. As a result, we choose to let the iteration process stop when either of the following two conditions are satisfied: (a) the residual norm of the state estimate is lower than a pre-set threshold $\beta_u \sqrt{p}$ for the first time, in light of (\ref{eq:residual_norm_target_ineq}); or (b) the maximum iteration number is reached. Condition (b) is introduced here to control the runtime of the iteration process. 

\begin{figure*} 
\centering
\includegraphics[width = 0.8\textwidth] {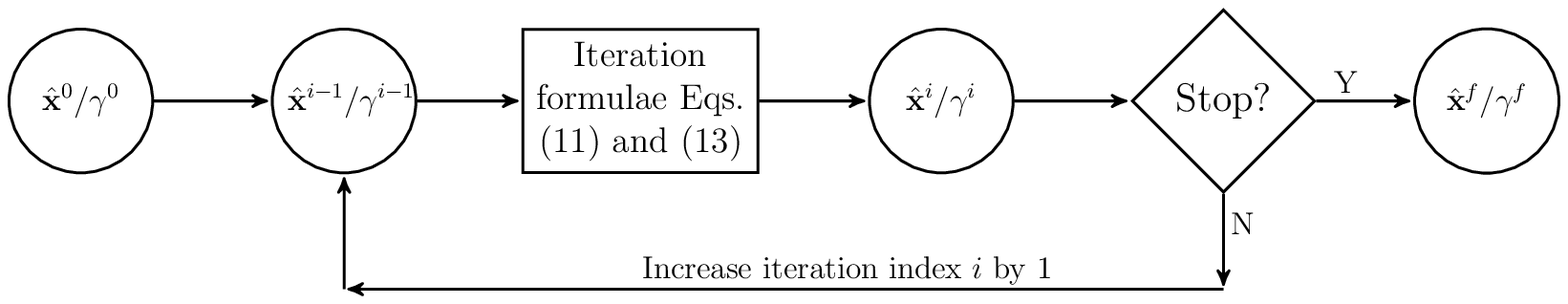}
\caption{\label{fig:iterative_framework} A schematic description of the iteration process.}
\end{figure*}

Figure \ref{fig:iterative_framework} provides a schematic outline of the iteration process. Given a pair of quantities $(\hat{\mathbf{x}}^{i-1},\gamma^{i-1})$ at the $(i-1)$-th iteration step, Eqs. (\ref{eq:iteration_formula}) and (\ref{eq:par_rule}) are applied to update them to  $(\hat{\mathbf{x}}^{i},\gamma^{i})$. Then, one checks if any of the above stopping conditions is met. If not, then Eqs. (\ref{eq:iteration_formula}) and (\ref{eq:par_rule}) are applied again to find new iterated values $(\hat{\mathbf{x}}^{i+1},\gamma^{i+1})$, and so on, until at least one of the stopping conditions is satisfied. At the end of the iteration process, one obtains the final state estimate $\hat{\mathbf{x}}^{f}$, which will then be used to construct the analysis ensemble as in the ETKF (to be further explained in the next sub-section).    

It is worth noting that Eq. (\ref{eq:iteration_formula}) is similar to the iteration formulae used in \citet{chen2013-levenberg,emerick2012ensemble} in the context of the ensemble smoother (ES, see \citealp{Evensen2000}), and in \citet{Stordal2013iterative} in the context of the iterative adaptive Gaussian mixture (AGM) filter \citep{Stordal-bridging-2010}.
In \citet{emerick2012ensemble}, a constraint, $\sum_i 1/\gamma_i = 1$, is imposed on the ES, implying that $\gamma_i \ge 1$. However, such a constraint may not guarantee that the data mismatch term in (\ref{eq:nonlinear_ls}) can be sufficiently reduced. For instance, one can design a sequence $\{\gamma^i\}$ with increasing values, e.g., $\gamma^{i+1} = \rho \gamma^{i+1}$ for some $\rho > 1$ such that $\{\gamma^i\}$ grows exponentially fast but still satisfies the constraint $\sum_i 1/\gamma_i = 1$. Then, if $\{\gamma^i\}$ becomes large enough, the gain matrix $\mathbf{G}^i$ in Eq. (\ref{eq:iteration_formula}) tends to zero exponentially fast such that the iteration formula in Eq. (\ref{subeq:iter_form}) would quickly make no significant change to the estimate $ \hat{\mathbf{x}}^{i}$ (results not shown). In \citet{chen2013-levenberg}, the formula is established through an approximation, by discarding certain model state terms during the iteration process of the standard Levenberg-Marquardt algorithm. The parameters $\gamma^i$ are determined through a way similar to the back-tracking line search method \citep{Nocedal-numerical} and may increase or decrease, depending on the circumstances. The convergence of the residual norms of the corresponding iteration process is, however, not clear yet. The iteration formula in \citet{Stordal2013iterative} is similar to those in \citet{chen2013-levenberg,emerick2012ensemble}, but is derived from the point of view of the Bayesian inversion theory. Under suitable conditions, asymptotic optimality can be achieved through the iteration formula in the sense of \citet{Stordal2013New}.               

\subsection{Implementation in the framework of the ETKF} \label{subsec:iter_implementation}   

In this section, we consider incorporating the proposed iteration process (Eq.(\ref{eq:iteration_formula})) into the ETKF. The resulting filter is thus referred to as the iterative ETKF with residual nudging (IETKF-RN) hereafter. The idea here is to use the final model state $\hat{\mathbf{x}}^{f}$ of the iteration process as the analysis, but introduce no modification to the analysis square root matrix of the normal ETKF. That is to say, given the same background ensemble, the normal ETKF and the IETKF-RN yield identical analysis square root matrices, but in general they may have different analyses. The analysis ensemble in the IETKF-RN is then produced based on $\hat{\mathbf{x}}^{f}$ and the associated square root matrix, in exactly the same way as in the normal ETKF. The choice of the identical analysis square root matrix in both filters is motivated by the observation that, in the Kalman filter, the covariance update formula is independent of the prior or posterior mean (but not vice versa, see, for example, \citealp[ch. 5]{Simon2006}).               
               
The remaining issues then involve specifying the following quantities in the iteration process: the covariance $\hat{\mathbf{C}}^b$ and the Jacobian matrix $\mathbf{J}^i$ in Eq. (\ref{subeq:iter_gain}), and the initial value $\gamma^0$ and the reduction factor $\rho^i$ in Eq. (\ref{eq:par_rule}).   

\subsubsection{Specifying the covariance $\hat{\mathbf{C}}^b$}  
To evaluate the gain matrix in Eq. (\ref{sub_eq: gain_mtx}), one needs to compute the matrix product, $\mathbf{J}^{i} \hat{\mathbf{C}}^b (\mathbf{J}^{i})^T$, which may be computationally expensive in large-scale problems. To alleviate this problem, we take advantage of the fact that, for the purpose of residual nudging, it is not mandatory to relate $\hat{\mathbf{C}}^b$ to the background sample error covariance $\hat{\mathbf{P}}^b$. As a result, in our implementation, $\hat{\mathbf{C}}^b$ is constructed based on a background matrix $\mathbf{B}^{lt}$ that is the ``climatological'' covariance of a model trajectory from a long model run (see Section \ref{sec:experiments}\ref{subsec:exp_setting} on how $\mathbf{B}^{lt}$ is obtained). $\hat{\mathbf{C}}^b$ remains constant and thus avoids re-evaluation over time. To further reduce the computational cost, we let $\hat{\mathbf{C}}^b$ be a diagonal matrix, whose diagonal elements correspond to those of $\mathbf{B}^{lt}$. By doing so, the diagonal elements of $\hat{\mathbf{C}}^b$ may, on average, have magnitudes close to those of $\hat{\mathbf{P}}^b$, but require less time or storage when evaluating the relevant terms in Eq. (\ref{subeq:iter_gain}). On the other hand, if the computational cost is affordable, then it may be more preferable to use the full matrix $\mathbf{B}^{lt}$, or its hybrid with $\hat{\mathbf{P}}^b$, which may improve the estimation accuracy of the filter (results not shown). 

\textbf{Remark:} If $\hat{\mathbf{C}}^b$ is taken as a hybrid of $\mathbf{B}^{lt}$ and $\hat{\mathbf{P}}^b$, then the IETKF-RN draws similarity to the hybrid EnKF-3DVar scheme \citep{Hamill-hybrid}, in which the analysis mean at a given time instant is also obtained by minimizing a certain cost function, and the corresponding analysis ensemble is generated in the same way as in a certain EnKF (e.g., the EnKF with perturbed observations). 
As a further extension, one may also consider incorporating into (\ref{eq:nonlinear_ls}) the observations in multiple assimilation cycles. In this case, the iterative framework is used as an (iterative) ensemble smoother (see, for example, \citealp{yang2012handling,bocquet2013joint,bocquet2013iterative}), and is similar to certain hybrid ensemble 4DVar schemes (see, for example, \citealp{liu2008ensemble}).              

\subsubsection{Evaluating the Jacobian matrix $\mathbf{J}^i$} 
If the derivative of the observation operator $\mathcal{H}$ is known, then $\mathbf{J}^i$ can be explicitly constructed. In certain situations, although it is possible for one to evaluate the function values of $\mathcal{H}$, the function form itself may be too complex or even unknown to the user\footnote{Examples may include, for instance, neural networks or certain commercial softwares.}. It is therefore challenging to obtain the analytic form of the Jacobian matrix. To this end, below we adopt a stochastic approximation method, called simultaneous perturbation stochastic approximation (SPSA, see, for example, \citealp{spall1992multivariate}), to approximate $\mathbf{J}^i$. The main reason for us to adopt the SPSA method is that it is a relatively simple approximation scheme. In real applications, however, one may replace the SPSA method by more accurate -- but possibly also more sophisticated and expensive -- approximation schemes.

In the SPSA method, to evaluate $\mathbf{J}^i$ around $\hat{\mathbf{x}}^{i}$, a random perturbation $\delta \mathbf{e} = (\delta e_1, \dotsb, \delta e_m)^T$ is first generated, where $\delta e_j$ ($j = 1,\dotsb,m$) takes the value $1$ or $-1$ with equal probability. Let $\delta \mathbf{p} =  \mathbf{S} \delta \mathbf{e}$, where $\mathbf{S}$ is an $m \times m$ matrix, e.g., a square root of $\hat{\mathbf{C}}^b$ as in our implementation. In addition, let the elements of $\delta \mathbf{p}$ be $\delta p_j$ ($j = 1,\dotsb,m$), and define a corresponding point-wise inverse vector $\delta \mathbf{p}_{inv} \equiv ((\delta p_1)^{-1}, \dotsb, (\delta p_m)^{-1})^T$. Then, we calculate the approximate Jacobian from
\begin{linenomath*}
\begin{equation} \label{eq:SPSA_appro}
\mathbf{J}^i \approx \dfrac{\mathcal{H}(\hat{\mathbf{x}}^{i} + \alpha \delta \mathbf{p} ) - \mathcal{H}(\hat{\mathbf{x}}^{i} - \alpha \delta \mathbf{p} )}{2 \alpha} \times \delta \mathbf{p}_{inv} \, , 
\end{equation}
\end{linenomath*}      
where $\alpha$ is a scaling factor. Eq. (\ref{eq:SPSA_appro}) can be considered as a stochastic implementation of the finite difference scheme for Jacobian approximation. From this point of view, $\alpha$ may take some relatively small value, e.g., $\alpha = 10^{-3}$ as in our implementation.


\subsubsection{Updating the parameter $\gamma^i$}\label{subsubsec:para_rule}

The initial value $\gamma^0$ is chosen in a way such that relatively small changes are introduced to $\hat{\mathbf{x}}^{0}$, in light of the requirement of linearizing $\mathcal{H}$ (through SPSA) in our implementation. To this end, in Eq. (\ref{subeq:iter_gain}), the terms $\mathbf{J}^{0} \hat{\mathbf{C}}^b (\mathbf{J}^{0})^T$ and $\gamma^0 \, \mathbf{R}$ are made comparable, by letting $\gamma^0 = \text{trace}(\mathbf{J}^{0} \hat{\mathbf{C}}^b (\mathbf{J}^{0})^T) / \text{trace}(\mathbf{R})$ \footnote{If necessary, one may choose a larger value for $\gamma^0$ (meaning a smaller step size in Eq. (\ref{eq:iteration_formula})), in order for a more accurate first order Taylor approximation in (\ref{eq:linearized_cost_func}). A consequence of such a choice, however, is that more iteration steps may be needed to reduce the residual norm by the same amount.}, where $\text{trace}(\bullet)$ denotes the trace of a matrix.  

The deterministic inverse problem theory (see, for example, \citealp[ch. 11]{Engl2000-regularization}) suggests that any parameter rule satisfying (\ref{eq:par_rule}) is sufficient for the purpose of residual nudging. In our implementation, however, $\mathbf{J}^{i} \hat{\mathbf{C}}^b (\mathbf{J}^{i})^T$ may be singular. Therefore, if $\gamma^i$ approaches zero too fast during the iteration (e.g., by letting $\gamma^{i+1} = \rho \, \gamma^i$ for a constant $\rho \in (0,1)$), then one would quickly encounter numerical problems when inverting $\mathbf{J}^{i} \hat{\mathbf{C}}^b (\mathbf{J}^{i})^T + \gamma^i \mathbf{R}$ in Eq. (\ref{subeq:iter_gain}). To avoid this problem, in our implementation, we let $\gamma^{i+1} = \gamma^i \, e^{-1/i}$ ($i = 1, 2, \dotsb$). The reduction factor $e^{-1/i}$ approaches 1 as the iteration index increases, while the parameters $\gamma^i$ and $\gamma^{i+1} $ still satisfy (\ref{eq:par_rule}). In the same spirit, one may adopt similar parameter rules, e.g., $\gamma^{i+1} = \gamma^i \, (1 - 1/(i+1))$ ($i = 1, 2, \dotsb$), which also worked well in our experiments (results not shown).   

\section{Experiments} \label{sec:experiments}

\subsection{Experimental settings} \label{subsec:exp_setting}
The $40$-dimensional L96 model \citep{Lorenz-optimal} is adopted as the testbed. The governing equations of the L96 model are given by
\begin{linenomath*}
\begin{equation} \label{eq:LE98}
\frac{dx_i}{dt} = \left( x_{i+1} - x_{i-2} \right) x_{i-1} - x_i + F, \, i=1, \dotsb, 40.
\end{equation}
\end{linenomath*}
We define $x_{-1}=x_{39}$, $x_{0}=x_{40}$ and $x_{41}=x_{1}$ in Eq.~(\ref{eq:LE98}).

The L96 model is integrated by the fourth-order Runge-Kutta method with a constant integration step of $0.05$. In many of the experiments below, the following default settings are adopted unless otherwise stated: the L96 model is integrated from time $0$ to $75$ (Section \ref{sec:experiments}\ref{subsec:experiment_results}-\ref{subsubsec:experiment_comparison}) or $525$ (Section \ref{sec:experiments}\ref{subsec:experiment_results}-\ref{subsubsec:experiment_stability}) with the forcing term $F=8$. To avoid the transition effect, the trajectory between $0$ and $25$ is discarded, and the rest ($1000$ and $10000$ integration steps in Sections \ref{sec:experiments}\ref{subsec:experiment_results}-\ref{subsubsec:experiment_comparison} and \ref{sec:experiments}\ref{subsec:experiment_results}-\ref{subsubsec:experiment_stability}, respectively) is used as the truth in data assimilation. For convenience, we relabel the time step at 25.05 as step 1. The synthetic observation $\mathbf{y}_k$ is obtained by measuring the odd number elements ($x_{k,1},x_{k,3},\dotsb$) of the state vector $\mathbf{x}_k = [x_{k,1},x_{k,2},\dotsb,x_{k,40}]^T$ every $4$ time steps ($k = 4, 8, 12, \dotsb$), in which the observation operator is given by $\mathcal{H} (\mathbf{x}_k) = (f(x_{k,1}),f(x_{k,3}),\dotsb,f(x_{k,39}))^T$, with $f(x) = x^3/5$ being a cubic function. The observation error is assumed to follow a normal distribution, $N(0,1)$, for each element in the observation vector. In some experiments, the forcing term $F$, the length of the assimilation time window, the frequency/density of the observations, the observation operator and so on, may be varied to investigate the sensitivities of the filter's performance to these factors.

To generate the initial background ensemble, we run the L96 model from $0$ to $5000$ (overall $100000$ integration steps), and compute the temporal mean $\mathbf{x}^{lt}$ and covariance $\mathbf{B}^{lt}$ of the trajectory. We then assume that the initial state vector follows the normal distribution $N(\mathbf{x}^{lt},\mathbf{B}^{lt})$, and draw a given number of samples as the initial background ensemble (which, of course, may not be the best possible way). $\mathbf{B}^{lt}$ is also used to construct the matrix $\hat{\mathbf{C}}^b$ in Eq. (\ref{eq:iteration_formula}), with $\hat{\mathbf{C}}^b = \text{diag}(\mathbf{B}^{lt})$ as previously mentioned, where $\text{diag}(\mathbf{B}^{lt})$ stands for the diagonal matrix whose diagonal elements are those of $\mathbf{B}^{lt}$. In the experiments, the stopping conditions of the iteration process in Eq. (\ref{eq:iteration_formula}) are either (a) when the residual norm is less than $2\sqrt{p}$, with $p$ being the observation size; or (b) when the iteration number reaches the maximum of 15000 iterations. In some cases the maximum iteration number may also change.

In all the experiments below, neither covariance inflation, nor covariance localization, is applied to the IETKF-RN. The former choice is because, in the presence of parameter $\gamma^i$ in Eq. (\ref{eq:iteration_formula}), conducting extra covariance inflation is equivalent to changing the initial value $\gamma^0$, which is investigated in an experiment below. With regard to localization, our experience suggests that, in some cases (for example, that with the default experimental settings at the beginning of this section and 20 ensemble members), conducting covariance localization may be beneficial for the IETKF-RN in the L96 model. In general, however, it is likely that the presence of covariance localization may alter the behaviour of IETKF-RN, in the sense that there is no guarantee any more that the iteration process (Eq. (\ref{eq:iteration_formula})), when equipped with covariance localization, moves along a residual-norm descent direction. Therefore, for our purpose, it appears more illustrative and conclusive for us to demonstrate only the performance of the IETKF-RN without localization.

\subsection{Experiment results} \label{subsec:experiment_results}

\subsubsection{A comparison study among some algorithms} \label{subsubsec:experiment_comparison}
A comparison study is first conducted to investigate the performance of the IETKF-RN relative to the following algorithms: the normal ETKF \citep{Bishop-adaptive,Wang-which}, the approximate Levenberg-Marquardt ensemble randomized maximum likelihood (LM-EnRML) method \citep{chen2013-levenberg}, and the iteration process of Eq. (\ref{eq:iteration_formula}) with $\gamma^i = 1~\forall~i$ fixed during the iteration process (for distinction, we call this algorithm ``IETKF-RN (constant $\gamma$)''). In the last algorithm, the iteration process aims to find a (local) minimum with respect to the cost function in Eq. (\ref{eq:nonlinear_ls}) with $\gamma=1$, which is essentially the same cost function adopted in, for example, the MLEF \citep{Zupanski-maximum}. In this sense, the ``IETKF-RN (constant $\gamma$)'' algorithm can be considered as an alternative to the MLEF, with one of the differences from the MLEF being in the chosen optimization algorithm: in the MLEF,  the conjugate gradient algorithm is adopted to minimize the cost function, while in the ``IETKF-RN (constant $\gamma$)'' algorithm, the Levenberg-Marquardt method is used instead. To show the necessity of using adaptive $\gamma$ values in certain circumstances, it would be desirable to conduct the comparison under the same conditions as far as possible. Therefore in what follows, we compare the IETKF-RN (with adaptive $\gamma$) with the ``IETKF-RN (constant $\gamma$)'', rather than directly with the MLEF.  

It is also worth commenting on a difference between the iteration processes of the IETKF-RN and the LM-EnRML. In the LM-EnRML, the terms $\hat{\mathbf{C}}^b (\mathbf{J}^{i})^T$ and $\mathbf{J}^{i} \hat{\mathbf{C}}^b (\mathbf{J}^{i})^T$ in Eq. (\ref{subeq:iter_gain}) are replaced by $\mathbf{S}_{x^i} (\mathbf{S}_{y^i})^T$ and $\mathbf{S}_{y^i} (\mathbf{S}_{y^i})^T$, respectively, where $\mathbf{S}_{x^i}$ and $\mathbf{S}_{y^i}$ are square root matrices of the sample covariances with respect to the ensemble $\mathbf{X}^i \equiv \{\mathbf{x}_1^i,\dotsb,\mathbf{x}_n^i\}$ at the $i$-th iteration and the corresponding projection $\mathbf{Y}^i \equiv \{\mathcal{H}(\mathbf{x}_1^i),\dotsb,\mathcal{H}(\mathbf{x}_n^i)\}$ onto the observation space, the same as those normally constructed in the EnKF (see, for example, \citealp{Luo-ensemble,Wang-which}). With such substitutions, though, there is no guarantee that the iteration moves toward a direction along which the residual norm is reduced. This point will be illustrated later. 

The normal ETKF is tested with the cubic observation function defined in the previous sub-section, with both covariance inflation and localization. In the experiments, we vary the inflation factor and half width of covariance localization within certain chosen ranges\footnote{Specifically the inflation factor $\delta \in \{1.05,~1.1,~1.15,\dotsb,1.30\}$, and the half width $l_c \in \{0.1, 0.3, 0.5, 0.7, 0.9\}$.}, and we observe that the normal ETKF ends up with large root mean squared errors (RMSEs) in all tested cases, suggesting that the normal ETKF has in fact diverged. Divergences of the EnKF have also been reported in other studies with nonlinear observations, see, for example, \citet{Jardak2010-comparison}.   

\begin{figure*} 
\vspace*{2mm}
\centering

\subfigure[Time series of residual norm]{ \label{subfig:rn_approxEnRMS}
\includegraphics[scale = 0.45]{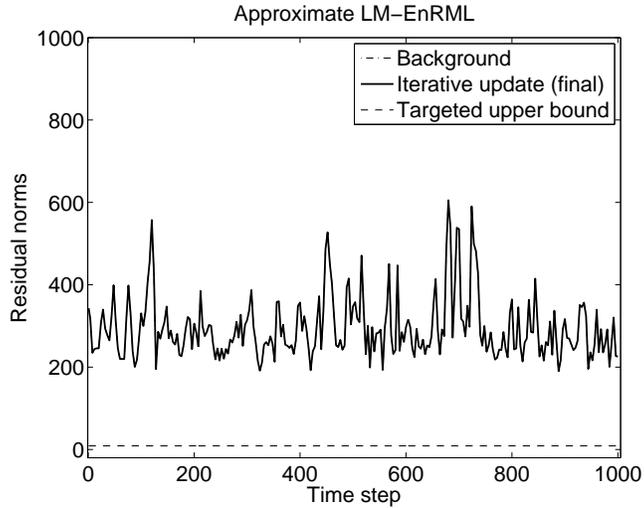}
}
\subfigure[Time series of RMSE]{ \label{subfig:rmse_approxEnRMS}
\includegraphics[scale = 0.45]{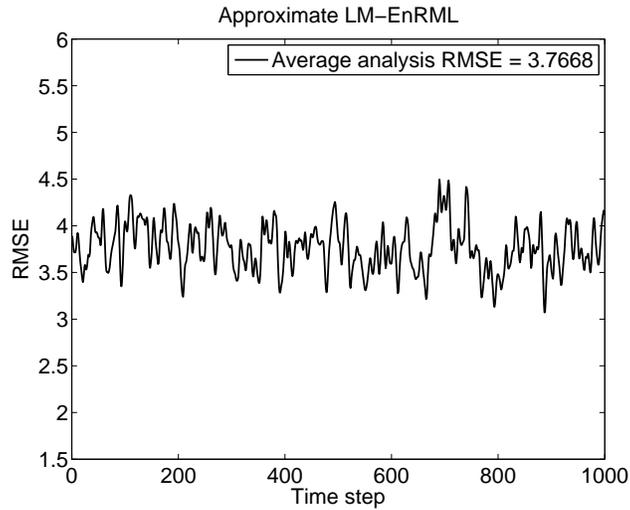}
}

\caption{\label{fig:approxEnRML} Time series of the residual norm and the corresponding RMSE of the approximate LM-EnRML method for the cubic observation operator. Note that in the upper panel, the time series of the background residual norm (dash-dotted) overlaps with that of the final analysis one (solid).}
\end{figure*}

Figure \ref{fig:approxEnRML} reports the time series of residual norms (upper panel) and the corresponding RMSEs (lower panel) obtained by applying the approximate LM-EnRM method with the same cubic observation function. The upper panel of Figure \ref{fig:approxEnRML} plots the background residual norm (dash-dotted line) and that of the final iterative estimate (also called the final analysis hereafter) of the iteration process (solid line), together with the targeted upper bound (dashed line), which is the threshold $\beta_u \sqrt{p} \approx 8.94$, with $\beta_u=2$ and $p= 20$ here. The time series of the final analysis residual norm overlaps with that of the background one in every assimilation cycle. They are thus indistinguishable in the figure. The reason for this result, in our opinion, is possibly that in this particular case, the approximate LM-EnRM method does not find an iterated estimate that is able to reduce the residual norm averaged over all ensemble members (a criterion used in \citealt{chen2013-levenberg} in order to update the estimate). As a result, following the parameter rule in \citet{chen2013-levenberg}, the $\gamma$ value in the iteration formula continues to increase and eventually results in a negligible gain matrix, such that the final analysis estimate is essentially almost the same as the background. Consequently, in this case, there is almost no residual norm reduction. Instead, the background and final analysis residual norms appear identical and stay away from the targeted upper bound. For reference, the time series of the RMSEs of the final analysis estimates is also plotted in the lower panel of Figure \ref{fig:approxEnRML}, with the corresponding time mean RMSE being about 3.77.

\begin{figure*} 
\vspace*{2mm}
\centering

\subfigure[Time series of residual norm]{ \label{subfig:rn_iETKF_cubic_fixedGamma}
\includegraphics[scale = 0.4]{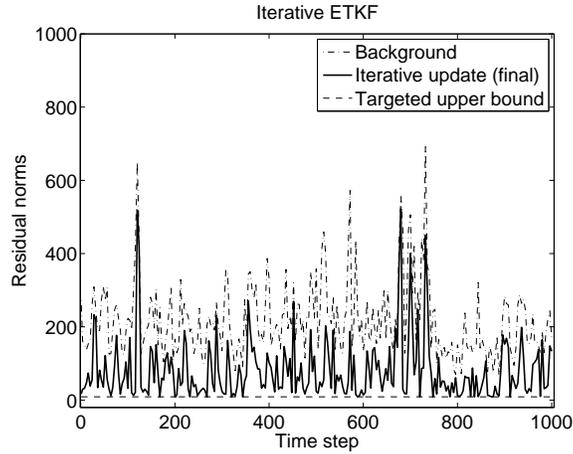}
}

\subfigure[Residual norm reduction of the iteration process at time step 500]{ \label{subfig:rmse_iETKF_cubic_iteration_fixedGamma}
\includegraphics[scale = 0.4]{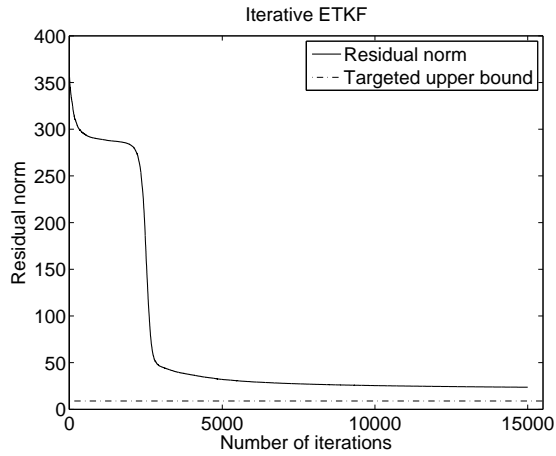}
}

\subfigure[Time series of RMSE]{ \label{subfig:rmse_iETKF_cubic_fixedGamma}
\includegraphics[scale = 0.4]{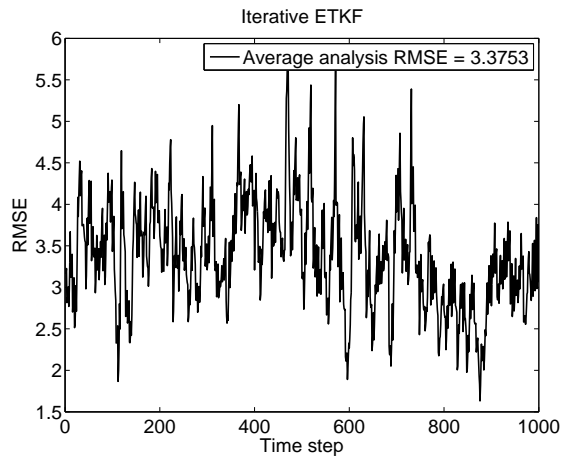}
}

\caption{\label{fig:iETKF_fixedGamma} IETKF-RN with fixed $\gamma=1$ applied to cubic observations. (a) Upper panel: time series of residual norms over the assimilation time window; (b) Middle panel: residual norm (solid) reduction of the iteration process at time step 500, an example used to illustrate how the residual norms of the iterative estimates are gradually reduced toward the targeted upper bound (dash-dotted); (c) Lower panel: time series of RMSEs of the final analysis estimates over the assimilation time window.}
\end{figure*}

Figure \ref{fig:iETKF_fixedGamma} plots the time series of the residual norms over the assimilation time window (upper panel); residual norm reduction of the iteration process at time step 500 (middle panel), an example that illustrates gradual residual norm reduction during the iteration process; and the time series of the corresponding RMSEs of the final estimates (lower panel), when the ``IETKF-RN (constant $\gamma$)'' algorithm is adopted to assimilate the cubic observations. Compared with Figure \ref{fig:approxEnRML}, it is clear that in the upper panel of Figure \ref{fig:iETKF_fixedGamma}, the residual norms of the final analysis estimates tend to be lower than the background ones in each assimilation cycle. In particular, in some cases, the final analysis residual norms approach, or even become slightly lower than, the pre-chosen upper bound of $8.94$, while the corresponding initial background residual norms are often larger than 100. As a consequence of residual norm reduction, the corresponding time mean RMSE in the lower panel reduces to 3.38, smaller than that in Figure \ref{fig:approxEnRML}. Also note that the time series of the residual norms (upper panel) appears spiky. This may be because the estimation errors at certain time instants are relatively large (although the corresponding final analysis residual norms may have reasonable magnitudes). Consequently, after model propagation, the resulting background ensembles may have relatively large residual norms. In addition, the iteration process at those particular time instants may converge slowly, or may be trapped around certain local optima, such that the final analysis residual norms are only slightly lower than the background ones (hence the spikes). This phenomenon is also found in other experiments later.

\begin{figure*} 
\vspace*{2mm}
\centering

\subfigure[Time series of residual norm]{ \label{subfig:rn_iETKF_cubic}
\includegraphics[scale = 0.4]{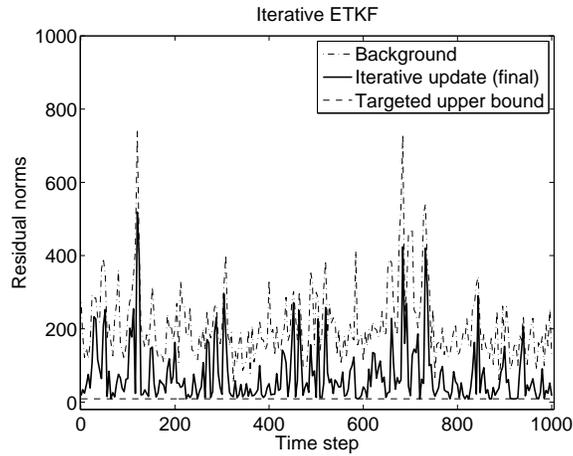}
}

\subfigure[Residual norm reduction of the iteration process at time step 500]{ \label{subfig:rmse_iETKF_cubic_iteration}
\includegraphics[scale = 0.4]{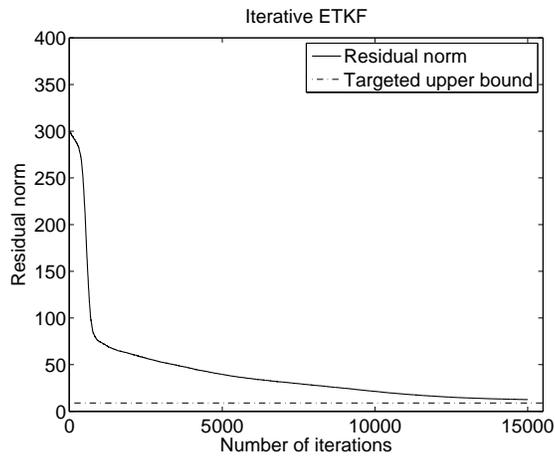}
}

\subfigure[Time series of RMSE]{ \label{subfig:rmse_iETKF_cubic}
\includegraphics[scale = 0.4]{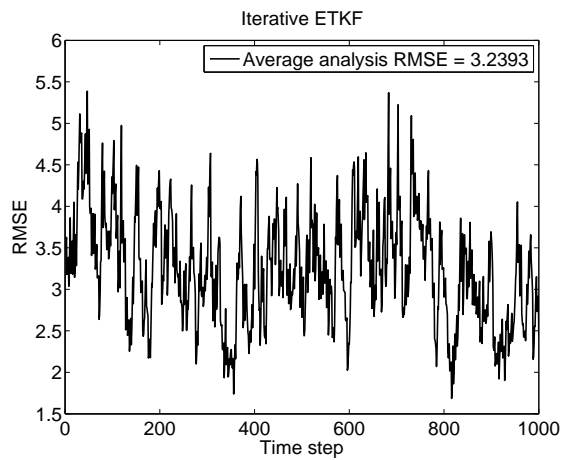}
}

\caption{\label{fig:iETKF_cubic} As in Figure \ref{fig:iETKF_fixedGamma}, except that now the IETKF-RN is associated with adaptive $\gamma$ during the iteration process.}
\end{figure*}

Similar results (see Figure \ref{fig:iETKF_cubic}) are also observed when the iteration process in Eq. (\ref{eq:iteration_formula}) is adopted, in conjunction with the adaptive parameter rule as described in Section \ref{sec:rn_nonlinear}\ref{subsec:iter_implementation}, to assimilate the cubic observations. One may see that the time mean RMSEs in Figs. \ref{fig:iETKF_fixedGamma} and \ref{fig:iETKF_cubic} are close to each other. In this sense, it appears acceptable in this case to simply take $\gamma^i = 1 $ for all $i$, instead of adopting the more sophisticated parameter rule in Section \ref{sec:rn_nonlinear}\ref{subsec:iter_implementation}. 

\begin{figure*} 
\vspace*{2mm}
\centering

\subfigure[Time series of residual norm]{ \label{subfig:rn_iETKF_exp}
\includegraphics[scale = 0.4]{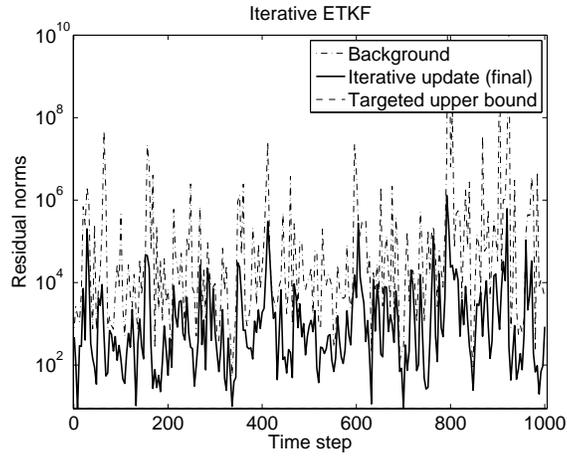}
}

\subfigure[Residual norm reduction of the iteration process at time step 500]{ \label{subfig:rmse_iETKF_exp_iteration}
\includegraphics[scale = 0.4]{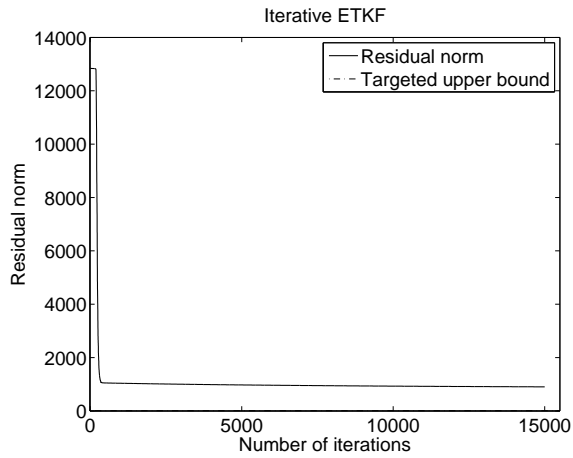}
}

\subfigure[Time series of RMSE]{ \label{subfig:rmse_iETKF_exp}
\includegraphics[scale = 0.4]{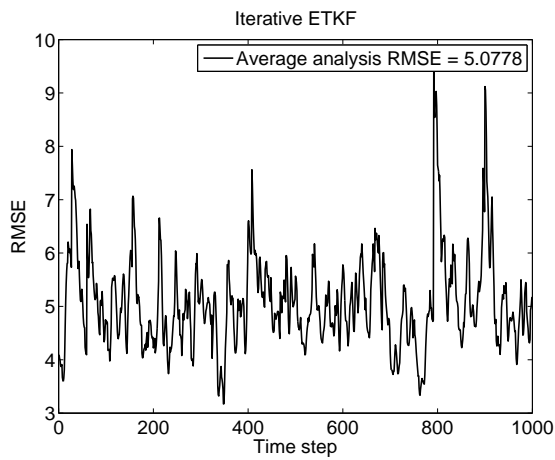}
}

\caption{\label{fig:iETKF_exp} As in Figure \ref{fig:iETKF_fixedGamma}, but now the IETKF-RN with adaptive $\gamma$ is applied to exponential observations.}
\end{figure*}

In what follows, though, we show with an additional example that the iteration process, when equipped with the adaptive parameter rule in Section \ref{sec:rn_nonlinear}\ref{subsec:iter_implementation}, tends to make the IETKF-RN more stable against filter divergence. To this end, we consider an exponential observation function $f(x) = e^{x^2/10}$ that is applied to the odd number elements (i.e., $x_1,x_3,\dotsb,x_{39}$) of the state vector. For such strongly nonlinear observations, the ``IETKF-RN (constant $\gamma$)'' algorithm diverges after 30 time steps. In contrast, the IETKF-RN with adaptive $\gamma^i$ appears to be more stable. As shown in the upper panel of Figure \ref{fig:iETKF_exp}, it still manages to reduce the background residual norm in all assimilation cycles. In fact, at a few time steps, the final analysis residual norms also approach the targeted upper bound. Compared with the cubic observation case (see, for example, Figure \ref{fig:iETKF_cubic}), however, the final analysis residual norms appear much larger in many assimilation cycles, due to the stronger nonlinearity in the exponential observation operator. 

The better stability of the IETKF-RN with adaptive $\gamma$, in comparison to the ``IETKF-RN (constant $\gamma$)'' algorithm in the case of exponential observations, may be understood from the optimization-theoretic point of view, when the iteration process in Eq.(\ref{eq:iteration_formula}) is interpreted as a gradient-based optimization algorithm. For this type of optimization algorithm, it is usually suggested to start with a relatively small step size, so that the linearization involved in the algorithms may remain roughly valid \citep{Nocedal-numerical}. In this regard, the IETKF-RN with adaptive $\gamma$ may appear to be more flexible (e.g., one may make the initial step size small enough by choosing a large enough value for $\gamma^0$), while there is no guarantee that the ``IETKF-RN (constant $\gamma$)'' algorithm may produce a small enough step size in general situations. 

\subsubsection{Stability of the IETKF-RN under various experimental settings} \label{subsubsec:experiment_stability}
Here we mainly focus on examining the stability of the IETKF-RN with adaptive $\gamma$ under various experimental settings. To this end, in the experiments below, we adopt assimilation time windows that are longer than those in the previous section. We note that the stability of the algorithm demonstrated below should be interpreted within the relevant experimental settings, and should not be taken for granted under different conditions (e.g., when with longer assimilation time windows).

Unless otherwise mentioned, in this section, the default experimental settings are as follows. The IETKF-RN is applied to assimilate cubic observations of the odd number state variables every 4 time steps, with the length of the assimilation time window being 10000 time steps. The variances of observation errors are 1. The IETKF-RN runs with 20 ensemble members and a maximum of 15000 iteration steps. 
 
In the first experiment, we examine the performance of the IETKF-RN with both linear and nonlinear observations (linear observations are obtained by applying $f(x) = x$ to specified state variables, plus certain observation errors).  For either linear or nonlinear observations, there are two observation scenarios: one with all 40 state variables being observed (the full observation scenario), and the other with only the odd number state variables being observed (the half observation scenario). In each observation scenario, we consider the following four ensemble sizes: 5, 10, 15 and 20. For each ensemble size, we also vary the frequency, in terms of the number $f_a$ of time steps, with which the observations are assimilated (i.e., the observations are assimilated every $f_a$ time steps). In the experiment, the variances of observation errors are 1, and $f_a$ is taken from the set $\{1, 2, 4:4:60\}$, where the notation $v_{i}:\delta v : v_f$ is used to denote an array of numbers that grows from the initial value $v_i$ to the final one $v_f$, with an even increment $\delta v$ each time. 

\begin{figure*} 
\vspace*{2mm}
\centering

\subfigure[Full (linear) observation scenario]{ \label{subfig:linObs_rmse_varying_obsOP_ensize_obsFreq_obsSkip-1}
\includegraphics[scale = 0.4]{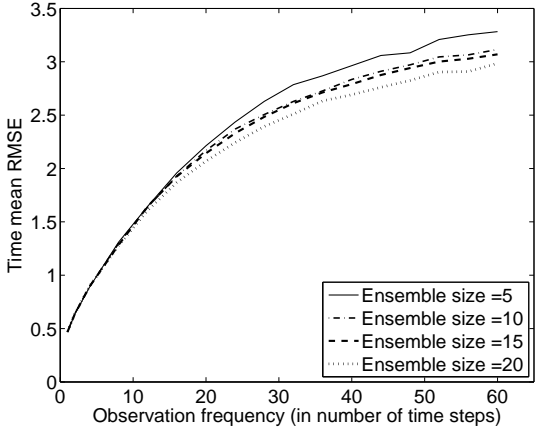}
} \hfill
\subfigure[Full (nonlinear) observation scenario]{ \label{subfig:rmse_varying_obsOP_ensize_full}
\includegraphics[scale = 0.4]{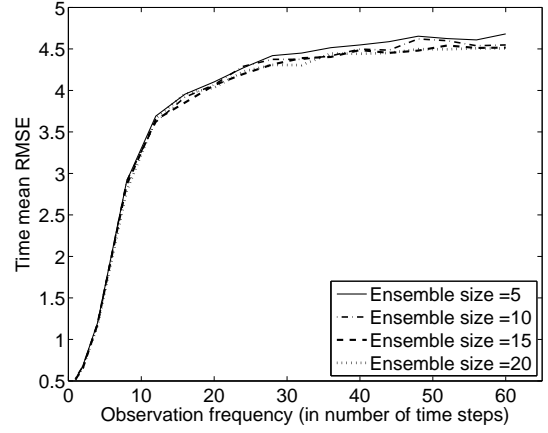}
} 

\subfigure[Half (linear) observation scenario]{ \label{linObs_rmse_varying_obsOP_ensize_obsFreq_obsSkip-2}
\includegraphics[scale = 0.4]{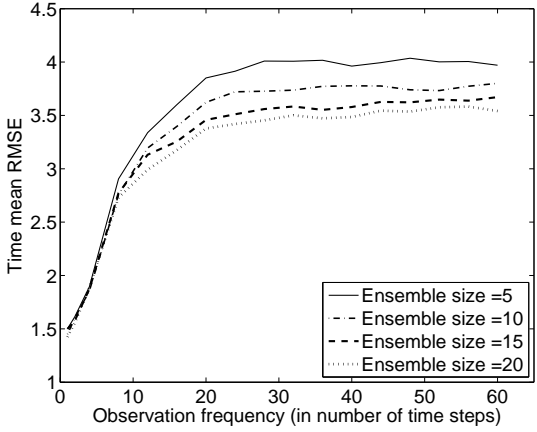}
}
\hfill
\subfigure[Half (nonlinear) observation scenario]{ \label{rmse_varying_obsOP_ensize_half}
\includegraphics[scale = 0.4]{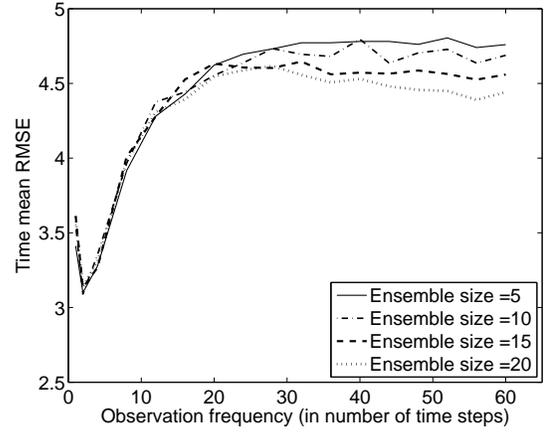}
} 

\caption{\label{fig:rmse_varying_obsOP_ensize_obsFreq_obsSkip2} Time mean RMSEs of the IETKF-RN as functions of the observation frequency (in number of time steps), for different ensemble sizes. The upper panels (a,b) are for the full observation scenario (p=40), and the lower ones (c,d) are for the half observation scenario (p=20). On the other hand, the left panels (a,c) are with linear observations, while the right ones (b,d) are with nonlinear observations. In all the above cases, the variances of observation errors are 1. Note that for the cases with linear observations, $\beta_u$ is set to 1. In addition, the corresponding Jacobians can be calculated exactly, so there is no need to apply SPSA to approximate them. }
\end{figure*}

Figure \ref{fig:rmse_varying_obsOP_ensize_obsFreq_obsSkip2} shows the time mean RMSEs (averaged over 10000 time steps) as functions of the ensemble size and the observation frequency, in the full and half observation scenarios, respectively. In the full observation scenario (upper panels) and the half observation scenario with linear observations (panel (c)), for each ensemble size, the corresponding time mean RMSE appears to be a monotonically increasing function of the number $f_a$ of time steps. On the other hand, when $f_a$ is relatively small, it appears that the time mean RMSEs of all ensemble sizes are close to each other. As $f_a$ increases, a larger ensemble size tends to yield a smaller time mean RMSE, although violations of this tendency may also be spotted in some cases of panel (b), possibly due to the sampling errors in the filter. In the half observation scenario with nonlinear observations (panel (d) of Figure \ref{fig:rmse_varying_obsOP_ensize_obsFreq_obsSkip2}), the behaviour of the IETKF-RN is similar to those in the other cases. There is, however, also a clear difference: instead of being a monotonically increasing function of $f_a$, the time mean RMSE in panel (d) exhibits V-shaped behaviour when $f_a$ is relatively small, achieving the lowest value at $f_a = 2$, rather than at $f_a = 1$ (possibly because the observations are over-fitted at $f_a = 1$). 

Overall, Figure \ref{fig:rmse_varying_obsOP_ensize_obsFreq_obsSkip2} indicates that, the time mean RMSEs with linear observations (panels (a,c)) tend to be lower than those with nonlinear observations (panels (b,d)), suggesting that the nonlinearity in the observations may deteriorate the performance of the filter. On the other hand, the time mean RMSEs in the full observation scenarios (panels (a,b)) tend to be lower than those in the half observation scenarios (panels (c,d)). The latter may be explained from the point of view of solving a (linear or nonlinear) equation. In the full observation scenarios, at each time step that has an incoming observation $\mathbf{y}^o$, the number of state variables is equal to the observation size $p$. Therefore (provided that it is solvable), the equation $\mathcal{H} (\mathbf{x}) = \mathbf{y}^o$ is well posed and has a unique solution. In this case, the smaller $f_a$ is, the more observations (hence constraints) there are. Consequently, there are fewer degrees of freedom in constructing the solution of the equation, and this tends to drive the state estimates toward the truth and yields relatively lower time mean RMSEs. In contrast, in the half observation scenario, the equation $\mathcal{H} (\mathbf{x}) = \mathbf{y}^o$ is under-determined (ill-posed) and in general has non-unique solutions. In this case, a smaller $f_a$ may tend to yield state estimates that better match the observations. However, similar to the issue of over-fitting observations in ill-posed problems, the smallest $f_a$ does not necessarily result in the lowest possible time mean RMSE, as shown in panel (d). 

As a side remark, we note that it is possible for one to further improve the performance of the IETKF-RN in Figure \ref{fig:rmse_varying_obsOP_ensize_obsFreq_obsSkip2} with other experimental settings. For instance, for the half observation scenario with linear observations (panel (c)), if one lets $\hat{\mathbf{C}}^b$ in Eq. (\ref{subeq:iter_gain}) be the sample covariance matrix of the background ensemble, and introduces both covariance inflation and localization to the filter, then in certain circumstances the time mean RMSE may be close to, or even lower than, 1 with relatively small $f_a$ values\footnote{For instance, when the inflation factor $\delta =0.08$, the half width $l_c = 0.1$ and $\beta_u = 1$, it is found that the time mean RMSEs of the IETKF-RN are around 0.50, 0.68 and 1.15, respectively, given $f_a = 1$, $2$ and $4$.}. However, for certain values of the half width of covariance localization, larger RMSEs or even filter divergence may also be spotted. We envision that this may be because in the presence of covariance localization, there is no guarantee that the iteration process (Eq. (\ref{eq:iteration_formula})) moves along a residual-norm descent direction. Therefore extra efforts are needed to take into account the impact of covariance localization on the search path of the iteration process, which will be investigated in the future.

\begin{figure*} 
\vspace*{2mm}
\centering

\subfigure[Time series of residual norm]{ \label{subfig:darn_output_longRun}
\includegraphics[scale = 0.4]{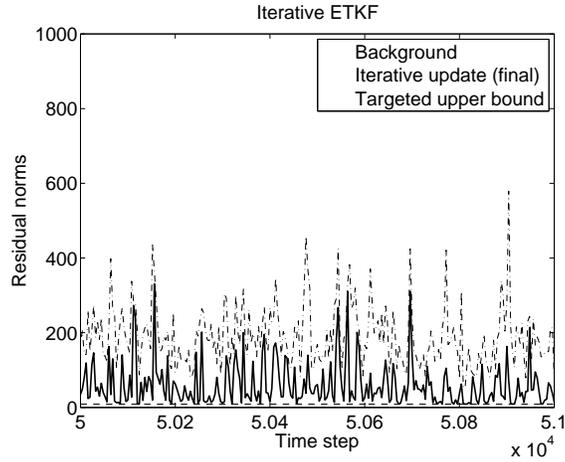}
}

\subfigure[Residual norm reduction of the iteration process at time step 50000]{ \label{subfig:darn_output_iteration_longRun}
\includegraphics[scale = 0.4]{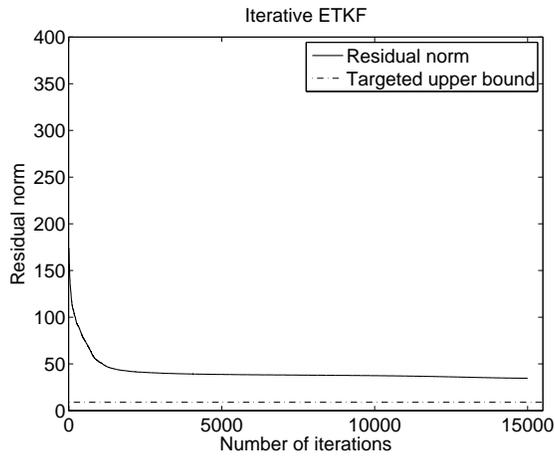}
}

\subfigure[Time series of RMSE]{ \label{subfig:darn_output_rmse_longRun}
\includegraphics[scale = 0.4]{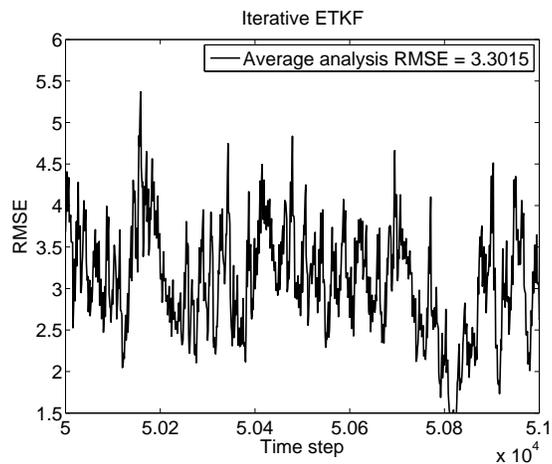}
}

\caption{\label{fig:iETKF_longRun} As in Figure \ref{fig:iETKF_cubic}, except that the length of the assimilation time window is now $100000$ time steps. For visualization, in the upper and lower panels we show only the time series between the time steps 50000 and 51000. The average time mean RMSE in the legend of the lower panel is, however, calculated with respect to the whole assimilation time window.}
\end{figure*}

We also follow \citet{sakov2012iterative} to test the IETKF-RN with a longer assimilation time window that consists of $100000$ time steps. Here the half (nonlinear) observation scenario is investigated, with the ensemble size being 20 and the observation frequency being every 4 time steps. Under these experimental settings, Figure \ref{fig:iETKF_longRun} shows that the IETKF-RN runs stably, and its time mean RMSE is around 3.30, close to the values in the corresponding panels of Figs. \ref{fig:iETKF_cubic} and \ref{fig:rmse_varying_obsOP_ensize_obsFreq_obsSkip2}. 

\begin{figure*} 
\vspace*{2mm}
\centering
\subfigure[Variances of observation errors are 0.01]{
	\includegraphics[scale = 0.45]{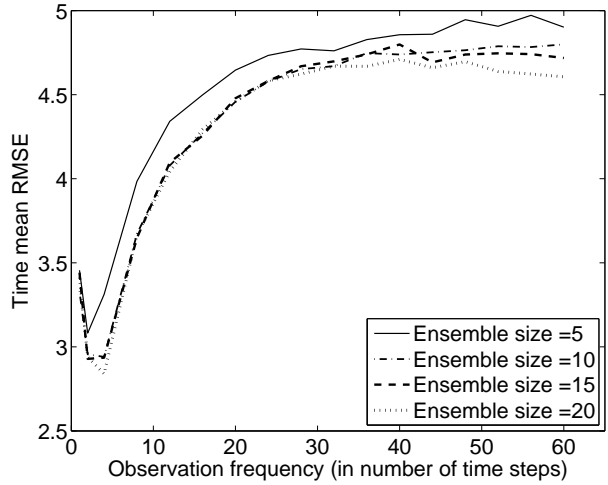}
}
\subfigure[Variances of observation errors are 10]{
	\includegraphics[scale = 0.45]{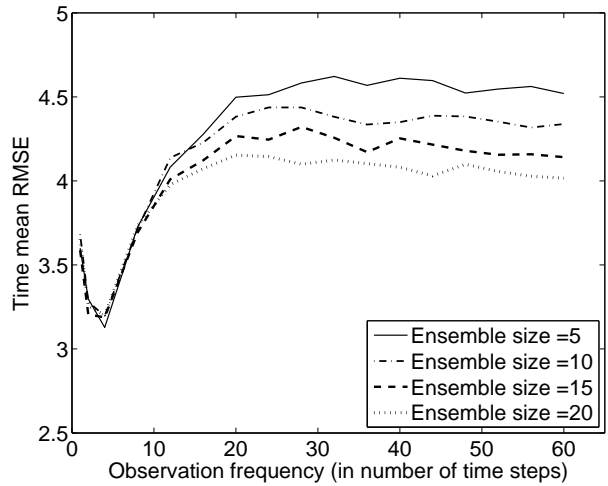}
}
\caption{\label{fig:rmse_varying_obsLvl_ensize_obsFreq_obsSkip2} Time mean RMSEs of the IETKF-RN as functions of the observation frequency (in number of time steps), for different ensemble sizes. Variances of observation errors in the upper panel are 0.01, and those in the lower panel are 10.}
\end{figure*}

Next, we test the performance of the IETKF-RN with different variances of observation errors. The experimental settings here are similar to those in panel (d) of Figure \ref{fig:rmse_varying_obsOP_ensize_obsFreq_obsSkip2}, except that the variances of observation errors are 0.01 and 10, respectively. As can been seen in Figure \ref{fig:rmse_varying_obsLvl_ensize_obsFreq_obsSkip2}, for these two variances, the IETKF-RN also runs stably for all tested ensemble sizes and observation frequencies. Comparing panel (d) of Figure \ref{fig:rmse_varying_obsOP_ensize_obsFreq_obsSkip2} and the panels of Figure \ref{fig:rmse_varying_obsLvl_ensize_obsFreq_obsSkip2}, the IETKF-RN exhibits similar behaviours in these cases. It also indicates that when $f_a$ is relatively small (say, $f_a = 4$), smaller variances tend to lead to lower time mean RMSEs; while when $f_a$ is relatively large (say, $f_a = 60$), the situation seems to be the opposite. Since both $f_a$ and the variances of observation errors affect the quality of the subsequent background ensembles, we conjecture that the above phenomenon occurs because for different combinations of $f_a$ and variances, different \textit{relative weights} are assigned to the background ensembles and the observations at the analysis steps. As a result, when $f_a$ is relatively large, smaller variances do not necessarily lead to lower time mean RMSEs. Similar results are also found in, for example, \citet[Figure 7]{Luo2014-efficient}.
   
\begin{figure*} 
\vspace*{2mm}
\centering

\includegraphics[scale = 0.45]{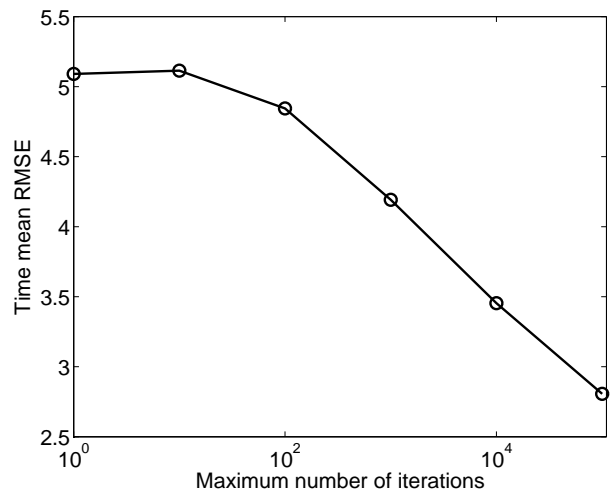}
\caption{\label{fig:iETKF_maxCount} Time mean RMSEs of the IETKF-RN as a function of the maximum number of iterations. With different orders of the magnitudes of the maximum iteration numbers, the logarithmic scale is used for the horizontal axis.}

\end{figure*}

We also investigate the effect of the maximum number of iterations on the performance of the IETKF-RN. The set of maximum numbers of iterations tested in the experiment is $\{1,10,100,1000,10000,100000\}$. Figure \ref{fig:iETKF_maxCount} shows the time mean RMSE of the IETKF-RN as a function of the maximum number of iterations. There, one can see that the time mean RMSE tends to decrease as the maximum number of iterations increases. 

\begin{figure*} 
\vspace*{2mm}
\centering

\includegraphics[scale = 0.8]{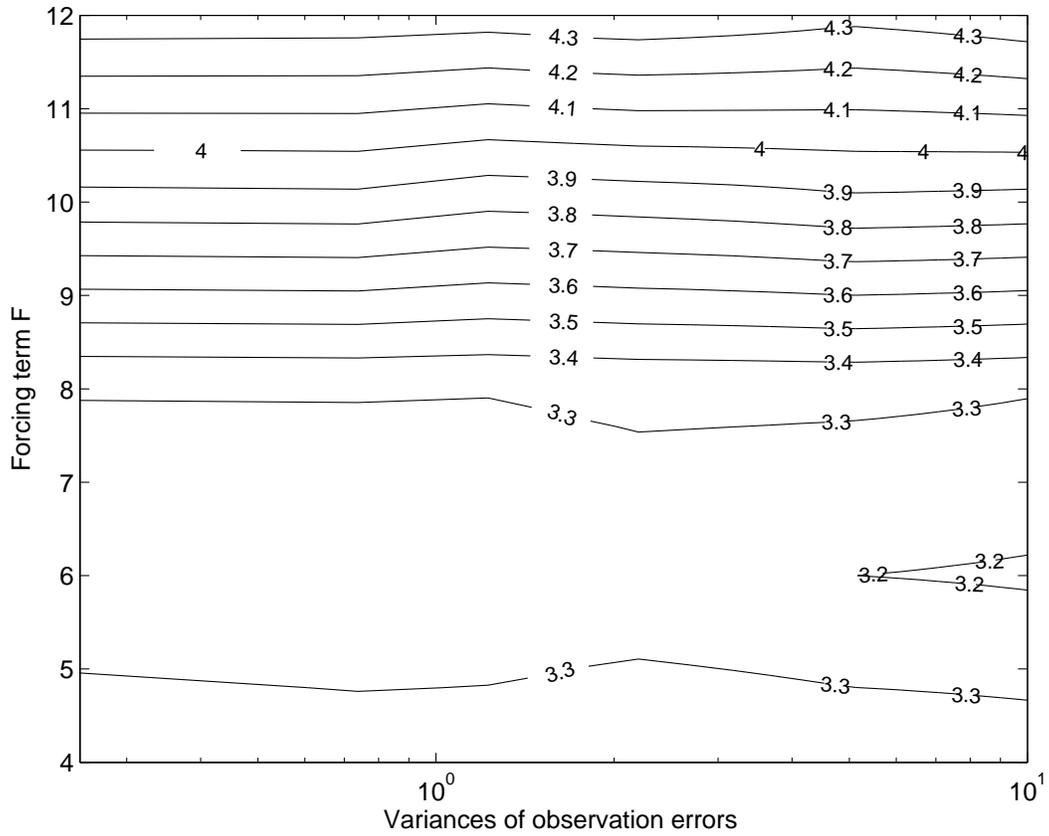}
\caption{\label{fig:rmse_F_vs_C} Time mean RMSE of the IETKF-RN as functions of the potentially mis-specified forcing term F and the variances of observation errors. For visualization, here the logarithmic scale is also used for the horizontal axis.}
\end{figure*}

Finally, we examine the impacts of (potentially) mis-specifying the forcing term $F$ in the L96 model and/or the variances of observation errors, on the performance of the IETKF-RN. In the experiment, the true forcing term $F$ is 8, and the true variances of observation errors are 1 for all elements of the observations. The tested $F$ values are taken from the set $\{4:2:12\}$, and the tested variances of observation errors are $\{0.25,0.5,1,2,5,10\}$. Figure \ref{fig:rmse_F_vs_C} reports the time mean RMSE as functions of the forcing term $F$ and the variances of observation errors. One can see that the time mean RMSE seems not very sensitive to the (potential) mis-specification of the variances of observation errors, possibly because with cubic observations, $\mathbf{J}^{i} \hat{\mathbf{C}}^b (\mathbf{J}^{i})^T$ in Eq. (\ref{eq:iteration_formula}) dominates $\gamma^i \mathbf{R}$ for a moderate initial $\gamma$ value. Therefore a mis-specification of $\mathbf{R}$ in the experiments may not have a substantial impact on the iteration process. If one increases the initial $\gamma$ value, or assimilates linear observations instead, then there can be more variations in the final estimation errors (results not shown).    

The performance of the IETKF-RN in Figure \ref{fig:rmse_F_vs_C}, on the other hand, does appear to be sensitive to the potential mis-specification of $F$. Interestingly, for all tested variances of observation errors, the filter's best performance is obtained at $F=6$, rather than $F=8$ \footnote{In the full observation scenario, however, the lowest time mean RMSE is indeed achieved at $F=8$ (results not shown).}. This suggests that, in certain situations, the filter might actually achieve better performance in the presence of certain suitable model errors, rather than with the perfect model. Similar observations are also reported in the literature, e.g., \citet{Gordon1993,Whitaker2012-evaluating}, in which it is found that introducing certain artificial model errors may in fact improve the performance of a data assimilation algorithm. Overall, Figure \ref{fig:rmse_F_vs_C} suggests that the IETKF-RN can also run stably even with substantial uncertainty in the system.  
   
\section{Conclusion} \label{sec:conclusion}

In this work, we introduced the concept of data assimilation with residual nudging. Based on the method derived in a previous study, we proposed an iterative filtering framework to handle nonlinear observations in the context of residual nudging. The proposed iteration process is related to the regularized Levenberg-Marquardt algorithm from inverse problem theory. Such an interpretation motivated us to implement the proposed algorithm with an adaptive coefficient $\gamma$. 

For demonstration, we implemented an iterative filter based on the ensemble transform Kalman filter (ETKF). Numerical results showed that the resulting iterative filter exhibited remarkable stability in handling nonlinear observations under various experimental settings, and that the filter achieved reasonable performance in terms of root mean squared errors.

For data assimilation in large-scale problems, it may not be realistic to conduct a large number of iterations, because of the limitation in computational resources. In this regard, one topic in our future research is to explore the possibility of enhancing the convergence rate of the iterative filter.
                      
\section*{Acknowledgment}

We would like to thank three anonymous reviewers for their constructive comments and suggestions. The first author would also like to thank the IRIS/CIPR cooperative research project ``Integrated Workflow and Realistic Geology'' which is funded by industry partners ConocoPhillips, Eni,  Petrobras, Statoil, and Total, as well as the Research Council of Norway (PETROMAKS) for financial support.

\bibliographystyle{ametsoc}
\bibliography{references}

\clearpage

\clearpage

\clearpage

\clearpage

\clearpage

\clearpage

\clearpage

\clearpage

\clearpage

\clearpage

\end{document}